\documentclass[conference,10pt,twocolumn,letter]{IEEEtran}

\IEEEoverridecommandlockouts 

 \usepackage{algorithm}
 \usepackage{algorithmic}
\usepackage{amsmath,amsfonts} 
\usepackage{amssymb, amsthm}
\usepackage{graphicx}
\usepackage{subfigure}
\usepackage{booktabs}
\usepackage{multirow}
\usepackage{mathrsfs}
\usepackage{cite}
\usepackage{units}
\usepackage{cancel}
\usepackage{upgreek}
\usepackage{color}
\DeclareGraphicsExtensions{.pdf,.jpeg,.png,.epbibdesks}
\usepackage{epstopdf}

\usepackage{microtype}

\renewcommand{\eqref}[1]{(\ref{#1})}
\newcommand{\secref}[1]{\mbox{Section~\ref{#1}}}

\newcommand{\abs}[1]{\ensuremath{\left|#1\right|}}

\newcommand{\setO}{\ensuremath{\mathcal{O}}}

\newcommand{\setX}{\ensuremath{\mathcal{X}}}

\newcommand{\bmb}{\ensuremath{\mathbf{b}}}

\newcommand{\bme}{\ensuremath{\mathbf{e}}}

\newcommand{\bmg}{\ensuremath{\mathbf{g}}}
\newcommand{\bmh}{\ensuremath{\mathbf{h}}}

\newcommand{\bmn}{\ensuremath{\mathbf{n}}}

\newcommand{\bmp}{\ensuremath{\mathbf{p}}}
\newcommand{\bmq}{\ensuremath{\mathbf{q}}}
\newcommand{\bmr}{\ensuremath{\mathbf{r}}}

\newcommand{\bmt}{\ensuremath{\mathbf{t}}}

\newcommand{\bmv}{\ensuremath{\mathbf{v}}}
\newcommand{\bmw}{\ensuremath{\mathbf{w}}}
\newcommand{\bmx}{\ensuremath{\mathbf{x}}}
\newcommand{\bmy}{\ensuremath{\mathbf{y}}}

\newcommand{\bmghat}{\ensuremath{\hat{\bmg}}}

\newcommand{\bmxhat}{\ensuremath{\hat{\bmx}}}

\newcommand{\bmgtilde}{\ensuremath{\tilde{\bmg}}}

\newcommand{\bmqtilde}{\ensuremath{\tilde{\bmq}}}

\newcommand{\bmvtilde}{\ensuremath{\tilde{\bmv}}}

\newcommand{\bmxtilde}{\ensuremath{\tilde{\bmx}}}

\newcommand{\bA}{\ensuremath{\mathbf{A}}}
\newcommand{\bB}{\ensuremath{\mathbf{B}}}
\newcommand{\bC}{\ensuremath{\mathbf{C}}}
\newcommand{\bD}{\ensuremath{\mathbf{D}}}
\newcommand{\bE}{\ensuremath{\mathbf{E}}}

\newcommand{\bG}{\ensuremath{\mathbf{G}}}
\newcommand{\bH}{\ensuremath{\mathbf{H}}}
\newcommand{\bI}{\ensuremath{\mathbf{I}}}

\newcommand{\bL}{\ensuremath{\mathbf{L}}}
\newcommand{\bM}{\ensuremath{\mathbf{M}}}

\newcommand{\bW}{\ensuremath{\mathbf{W}}}

\newcommand{\bZero}{\ensuremath{\mathbf{0}}}

\newcommand{\MT}{{\ensuremath{U}}}
\newcommand{\MR}{{\ensuremath{B}}}

\newcommand{\No}{\ensuremath{N_{\mathrm{0}}}}

\newcommand{\Es}{{\ensuremath{E_{\mathrm{s}}}}}



\begin{document}
\title{Conjugate Gradient-based Soft-Output Detection and Precoding in Massive MIMO Systems}

\author{
    \IEEEauthorblockN{Bei Yin$^\text{1}$, Michael Wu$^\text{1}$, Joseph R. Cavallaro$^\text{1}$, and Christoph Studer$^\text{2}$} \\ 
    \IEEEauthorblockA{$^\text{1}$Department of ECE, Rice University, Houston, TX; e-mail: \{by2,\,mbw2,\,cavallar\}@rice.edu}
   \IEEEauthorblockA{$^\text{2}$School~of ECE, Cornell University, Ithaca, NY; e-mail: studer@cornell.edu}
   \thanks{This work was supported in part by
   Texas Instruments, Xilinx, Samsung, Huawei, and by the US National Science Foundation under
   grants ECCS-1408370, CNS-1265332,  ECCS-1232274, and ECCS-1408006.} 
}


\maketitle

\begin{abstract}
Massive multiple-input multiple-output (MIMO) promises improved spectral efficiency, coverage, and range, compared to conventional (small-scale) MIMO wireless systems.
Unfortunately, these benefits come at the cost of significantly increased computational complexity, especially for systems with realistic antenna configurations. 
To reduce the complexity of data detection (in the uplink) and precoding (in the downlink) in massive MIMO systems, we propose to use conjugate gradient (CG) methods. 
While precoding using CG is rather straightforward, soft-output minimum mean-square error (MMSE) detection requires the computation of the post-equalization signal-to-interference-and-noise-ratio (SINR). To enable CG for soft-output detection, we propose a novel way of computing the SINR directly within the CG algorithm at low complexity.
We investigate the performance/complexity trade-offs associated with CG-based soft-output detection and precoding, and we compare it to existing exact and approximate methods. 
Our results reveal that the proposed algorithm is able to outperform existing methods for massive MIMO systems with realistic antenna configurations.
\end{abstract}


 
\section{Introduction} 
\label{sec:intro}
\subsection{Massive Multiple-Input Multiple-Output (MIMO)}

Massive (or large-scale) MIMO is an emerging technology to improve the spectral efficiency of existing (small-scale) MIMO wireless communication systems~\cite{Rusek2012,Marzetta2010}. 
The main idea is to equip the base station (BS) with hundreds of antennas that serve a (relatively) small number of users (in the orders of tens) simultaneously and in the same frequency band. 
Theoretical results for massive MIMO not only promise higher peak data rates, improved coverage, and longer range,  but also that simple, low-complexity, and energy-efficient detection and precoding algorithms are able to achieve optimum performance in the large-antenna limit, i.e., where the number of BS antennas approaches infinity~\cite{Rusek2012, Huh2011,Ngo2012,Marzetta2010}.

Unfortunately, systems with realistic antenna configurations (e.g., with a few hundred BS antennas or less) are far from the large-antenna limit. As a consequence, one still has to resort to computationally expensive detection and precoding algorithms to achieve near-optimal error-rate performance~\cite{hoydis2011massive}. 
As demonstrated in~\cite{Wu2012}, data detection (in the uplink) and precoding (in the downlink) are among the most challenging tasks in terms of computational complexity in realistic  systems. 
Hence, to reduce the computational complexity, linear and approximate data detection and precoding schemes, which rely on a truncated Neumann series expansion, have been proposed recently in~\cite{Wu2012,Yin2013,Wu2014,YWWDCS14b}. 
This approach requires (often significantly) lower computational complexity than that of an exact inversion while delivering near-optimal results for massive MIMO systems having a large ratio between BS and user antennas. 
However, this approximate detection approach suffers from a considerable error-rate performance loss if the ratio between BS antennas and user antennas is close to one~\cite{Wu2014}. 

\subsection{Contributions}

In this paper, we propose to use conjugate gradient (CG) methods for data detection and precoding in order to improve upon the Neumann series approach in~\cite{Wu2014,Wu2012,Yin2013,YWWDCS14b} for realistic massive MIMO systems.
While CG for precoding is rather straightforward, CG-based soft-output data detection necessitates the computation of the post-equalization signal-to-interference-and-noise-ratio (SINR), which typically requires the \emph{explicit inverse} of the channel matrix~\cite{Wu2014}. 
To avoid a matrix inversion altogether, we propose an exact and an approximate method to compute the SINR directly within the CG algorithm. We investigate the associated performance/complexity trade-offs and demonstrate that our CG-based detection and precoding method requires low computational complexity while outperforming exact and approximate linear methods in massive MIMO systems with realistic antenna configurations.

 
\section{Uplink and Downlink System Models}
\label{sec:system}

We consider the uplink and downlink of an orthogonal frequency-division multiplexing (OFDM)-based massive MIMO system where the BS is equipped with $\MR$ antennas and communicates with $\MT \leq \MR$  single-antenna users. 

\subsection{Uplink System Model and Soft-Output MMSE Detection}
\label{sec:uplinksystem}

In the uplink, the information bits for each user are encoded and mapped onto constellation points in the set $\setO$.   The modulated symbols are  transformed into the time domain using an inverse discrete Fourier transform (DFT). Each user then transmits the time-domain  signals over  the wireless channel. 
 
At the BS, the received signals are converted into frequency domain using DFTs. By omitting the subcarrier index, the equivalent {input-output} relation of the frequency-domain uplink channel for each subcarrier can be modeled as 
\begin{align} \label{eq:uplinkmodel}
\mathbf{y}_u=\mathbf{H}_u\mathbf{x}+\mathbf{n}_u,
\end{align}
where  $\bmx\in\setO^{\MT}$ is the transmit  vector with  modulated symbols from all users, and 
$\mathbf{y}_u\in\mathbb{C}^\MR$ is the receive-vector. 
The uplink channel matrix is given by $\bH_u\in\mathbb{C}^{\MR\times\MT}$, where we assume that each entry is generated from the WINNER-\mbox{Phase-2} channel model~\cite{winner2}. In what follows, we assume the channel matrix~$\bH_u$ to be perfectly known at the BS. 
Each element of noise vector $\bmn_u\in\mathbb{C}^{\MR}$ in \eqref{eq:uplinkmodel} is assumed to be i.i.d.\ circularly-symmetric complex Gaussian with variance~$\No$ per complex entry.
The per-user uplink transmit power for user $i$ is defined as $\mathbb{E}\{\abs{x_i}^2\}=\Es$ and the (average) uplink SNR is given by $UE_s/\No$.

In the remainder of the paper, we focus on soft-output MMSE detection, which achieves near-optimal performance in massive MIMO systems with large BS to user antenna ratios at manageable complexity~\cite{Wu2014}. 
Soft-output data detection requires the following well-known MMSE equalization matrix~\cite{Paulraj2008}: 
\begin{align} \label{eq:eqweight}
\bW=\left(\mathbf{H}^H_u\mathbf{H}_u + \varrho_u^{-1} \mathbf{I}_U\right)^{\!-1}\mathbf{H}^H_u,
\end{align}
where $\varrho_u = \Es/\No$ is the transmit SNR. 
To compute soft-output information in the form of log-likelihood ratio (LLR) values, we first compute estimates of the transmit symbol as 
\begin{align} \label{eq:detectionsignal}
\bmxhat=\bW\bmy.
\end{align}
By modeling the transmit symbol of user $i$ as $\hat{x}_i=\mu_ix_i+z_i$, where $\mu_i=\bmw_i^H\bmh_i$ is the equalized channel gain and $z_i=\sum_{j,j\neq i} \bmw_i^H\bmh_jx_j+\bmw_i^H\bmn$ models noise-plus-interference with variance $\nu_i^2=\mathbb{E}\{|z_i|^2\}$, we can compute the max-log approximated LLR of bit $b$ for user $i$ as follows~\cite{Studer2011}:
\begin{align} \label{eq:llr}
L_{i,b} = \rho_i \!\left(\min_{a\in\setX_b^{0}}\left|{\frac{\hat{x}_i}{\mu_i}}-a\right|^2\! \!-\!  \min_{a'\in\setX_b^{1}}\left|{\frac{\hat{x}_i}{\mu_i}}-a'\right|^2 \right)\!.
\end{align}
Here, $\rho_i=\mu_i^2/{\nu_i^2}$ is the post-equalization SINR, and the sets $\setX_b^{0}$  and $\setX_b^{1}$ contain the constellation symbols where bit $b$ of the symbol in $\setO$ equals $0$ and $1$, respectively. 

We emphasize that the method proposed in \secref{sec:algo} directly computes $\hat\bmx$ in \eqref{eq:detectionsignal} as well as the equalized channel gains~$\mu_i$, $\forall i$, and the post-equalization SINRs~$\rho_i$, $\forall i$, which avoids an explicit computation of \eqref{eq:eqweight} and \eqref{eq:detectionsignal}; this, in turn, significantly reduces the computational complexity of soft-output data detection in massive MIMO systems. 

\subsection{Downlink System Model and Precoding}
In the downlink, the BS encodes the bit streams for each user in the frequency domain. The encoded bits are then mapped to constellation points in $\setO$. The transmit vector $\bmt\in\setO^{\MT}$ containing the modulated symbols for all $\MT$ users is then processed using the following linear precoder:
\begin{align} \label{eq:precoding}
\mathbf{q}=\mathbf{P}\mathbf{t}.
\end{align}
Here, $\mathbf{P}$ is a $\MR \times \MT$  precoding matrix and $\mathbf{q}\in\mathbb{C}^\MR$ the  precoded vector.
To maximize the SINR at the receiver (and to mitigate inter-user interference), we deploy linear MMSE precoding, which is  defined as~\cite{Peel2005} 
\begin{align} \label{eq:precodingmatrix}
\mathbf{P}={\mathbf{H}_d^H}(\mathbf{H}_d\mathbf{H}_d^H + \varrho^{-1}_d\mathbf{I}_U)^{-1}.
\end{align}
Here, the downlink channel $\bH_d\in\mathbb{C}^{\MT\times\MR}$ satisfies $\bH_d=\bH_u^H$ due to reciprocity and $\varrho_d$ is the downlink SNR (as for the uplink).
We assume the downlink channel to be known and generated using the  WINNER-\mbox{Phase-2} channel model~\cite{winner2}. 
Note that prior to downlink transmission, we normalize the transmit power of the precoded vector as
$\mathbf{s}={\mathbf{q}}/{\|\mathbf{q}\|_2}$,
where $\mathbf{s}\in\mathbb{C}^{\MR}$ is the transmit vector normalized to unit power. 
The precoded and normalized vector $\mathbf{s}$ is then converted into the time domain and transmitted to the users.
The frequency-domain equivalent {input-output} relation of the downlink channel is  
$\bmy_d=\mathbf{H}_d\mathbf{s}+\bmn_d$,
where $\bmn_d\in\mathbb{C}^\MT$ models noise and $\bmy_d\in\mathbb{C}^\MT$ contains the receive symbols for each user. 

We emphasize that our precoding method proposed next directly computes the precoded vector $\bmq$ in \eqref{eq:precoding}, without explicitly forming the precoding matrix \eqref{eq:precodingmatrix}; this approach (often significantly) reduces  computational complexity of precoding.

\section{CG-Based Data Detection and Precoding}
\label{sec:algo}

Virtually all existing linear soft-output detection methods as well as some precoders \emph{explicitly} compute the MMSE equalization matrix~\eqref{eq:eqweight} or precoding matrix \eqref{eq:precodingmatrix}, which requires the inverse of a $\MT\times\MT$ matrix.
This matrix inversion incurs significant computational complexity, especially for a large number of user antennas~$\MT$. 
We now propose a novel approach to low-complexity soft-output data detection (Section~\ref{sec:cgdetection}) and precoding (Section~\ref{sec:cgprecoder}) for massive MIMO systems, which avoids such an explicit matrix inversion altogether. 
 
\subsection{Conjugate Gradient (CG) Basics}

CG is an efficient iterative method to solve systems of linear equations~\cite{Hestenes1952}. Specifically, CG solves problems of the form 
\begin{align} \label{eq:cgbasics}
\bmghat = \underset{\bmgtilde\in\mathbb{C}^\MT}{\operatorname{arg\,min}} \,  \|\bmb - \bA\bmgtilde\|,
\end{align} 
where $\bA\in\mathbb{C}^{\MT\times\MT}$ is a positive definite matrix. In contrary to direct methods that compute $\bmghat=\bA^{-1}\bmb$, CG iteratively computes the solution $\bmghat$ with each iteration requiring low computational complexity. One of the key advantages of CG is the fact that CG converges at $U$ iterations and the iterative procedure can be terminated early while still obtaining a solution close to the exact result $\bmghat$; this leads to (often significantly) lower computational complexity instead of directly inverting the matrix~$\bA$. 

Since \eqref{eq:cgbasics} can be used to compute the solutions to \eqref{eq:detectionsignal} and~\eqref{eq:precoding} with appropriate $\bmb$ and $\bA$, CG is---at least in principle---suitable for low-complexity linear data detection and precoding. 
However, the key disadvantage of CG is that it does not provide the post-equalization SINR information, which is necessary to compute LLR values \eqref{eq:llr}. To use CG for soft-output data detection, we  propose a novel  method that computes the necessary SINR information directly within CG. 

\subsection{CG-D: CG-Based Soft-Output Data Detection}  
\label{sec:cgdetection}

As mentioned above, the solution to \eqref{eq:detectionsignal} can be computed by solving the following optimization problem~\cite{Bulirsch91}:
\begin{align} \label{eq:CGuplink}
\bmxhat = \underset{\bmxtilde\in\mathbb{C}^\MT}{\operatorname{arg\,min}} \,  \|\bH_u^H\bmy - \bA\bmxtilde\|, 
\end{align}
where $\bA = \bH^H_u\bH_u+\varrho_u^{-1}\bI_\MT$ is the regularized uplink Gram matrix.
The solution $\bmxhat$ to \eqref{eq:CGuplink} can be computed (or approximated) efficiently using CG \cite{Bulirsch91}. 

Algorithm \ref{alg:CG} summarizes our CG-based approach for soft-output data detection. The base algorithm of our CG method follows that in~\cite{Bulirsch91}.
On line 5, we first compute the matched filter vector $\bH_u^H\bmy$ and compute the regularized Gram matrix~$\bA$. We then initialize the vectors $\bmv_0$,  $\bmr_0$,  and $\bmp_0$ used in the CG procedure. 
On lines 9--14, we iteratively compute~$\bmv_k$, which will be our MMSE estimate $\bmxhat_K$ after $K$ iterations (see line~19). Here, the intermediate results~$\bmv_k$,  $\bmr_k$, and $\bmp_k$ are computed recursively from $\bmv_{k-1}$,  $\bmr_{k-1}$,  and $\bmp_{k-1}$. The output of this CG method in the $K^\text{th}$ iteration, $\bmxhat_K=\bmv_K$, is an estimate for $\bmxhat$ in~\eqref{eq:detectionsignal} if the procedure is terminated before reaching $\MT$ iterations. Since CG is an exact method~\cite{Hestenes1952,Bulirsch91}, Algorithm~\ref{alg:CG} delivers the exact solution to \eqref{eq:detectionsignal} if $K=\MT$. 

We emphasize that CG methods \emph{do not} provide the necessary post-equalization SINR information, such as $\rho_i$ as well as $\mu_i$, which are required to compute the LLR values in \eqref{eq:llr}. Hence, regular CG is not suitable for soft-output detection.\footnote{Hard-output data detection would be straightforward, but typically entails a significant error-rate performance loss in coded communication systems.} 
We next propose two novel methods to compute $\mu_{i|k}$ and $\rho_{i|k}$ directly within each iteration $k$ of CG. The first method is exact, i.e., computes  $\mu_i$ and $\rho_i$ after $K=\MT$ iterations; the second method approximates both quantities at low computational complexity. 

\setlength{\textfloatsep}{8pt}
\begin{algorithm}[t]
\caption{CG for soft-output MMSE detection \& precoding\label{alg:CG}}
\begin{algorithmic}[1]
\STATE \textbf{input:}
\STATE $\quad\mathbf{H}_u$ and $\mathbf{y}$ \COMMENT{detection}
\STATE $\quad\mathbf{H}_d$ and $\mathbf{t}$ \COMMENT{precoding}
\STATE $\textbf{initialization:}$
\STATE $\quad\bmb=\bH^H_u\bmy$ and $\bA = \bH^H_u\bH_u+\varrho_u^{-1}\bI_U$ \COMMENT{detection}
\STATE  $\quad\bmb=\bmt$ and $\bA = \bH_d\bH_d^H+\varrho_d^{-1}\bI_U$ \COMMENT{precoding}
\STATE  $\quad\mathbf{v}_0=\mathbf{0}$, $\mathbf{r}_0=\bmb$, and  $\mathbf{p}_0=\mathbf{r}_0$
\FOR{$k = 1,\ldots, K$}
\STATE $\bme_{k-1}=\bA\bmp_{k-1}$
\STATE $\alpha_k=\|\bmr_{k-1}\|^2/(\bmp_{k-1}^H\bme_{k-1})$
\STATE $\bmv_k=\bmv_{k-1}+\alpha_k\bmp_{k-1}$
\STATE $\bmr_k=\bmr_{k-1}-\alpha_k\bme_{k-1}$
\STATE $\beta_k=\|\bmr_{k}\|^2/\|\bmr_{k-1}\|^2$
\STATE $\bmp_k=\bmr_{k}+\beta_k\bmp_{k-1}$
\STATE {compute} $\mu_{i|k}$, $\forall i$, as in \eqref{eq:mu} \COMMENT{detection}
\STATE {compute} $\rho_{i|k}$, $\forall i$, as in \eqref{eq:rho}  \COMMENT{detection}
\ENDFOR
\STATE \textbf{output:}
\STATE $\quad\bmxhat_K=\bmv_K$, $\mu_{i|K}$, $\forall i$, and $\rho_{i|K}$, $\forall i$, \COMMENT{detection}
\STATE $\quad\bmq_K=\bH^H_d\bmv_K$ \COMMENT{precoding}
\end{algorithmic}
\end{algorithm}


\subsection{SINR Computation Methods}  
\label{sec:SINRtrackers}

\subsubsection{Exact SINR Computation}

Although the equalized vector~$\bmxhat_{k}$ is computed in an iterative fashion (see Algorithm~\ref{alg:CG}), it can be computed  from the received vector $\bmy$ with a CG-equivalent equalization matrix that depends on the iteration index $k$. 
By defining the CG-equivalent equalization matrix as  $\bL_{k}\bH_u^{H}$, the $k^\text{th}$ estimate $\bmxhat_{k}$ can be computed as
\begin{align} \label{eq:eqmatrix}
\bmxhat_{k}=\bL_{k}\bH^{H}_u\bmy.
\end{align}
If the matrix $\bL_{k}$ is known, then the intermediate quantities $\mu_{i|k}$ and $\nu_{i|k}$ can be computed as detailed in Section~\ref{sec:uplinksystem}. 
We now propose a method to exactly compute $\bL_{k}$ in each iteration~$k$, which enables us to extract  $\mu_{i|k}$ and $\nu_{i|k}$ on-the-fly.

We start by inserting $\bme_{k-1}=\bA\bmp_{k-1}$ (from line~9) to $\bme_{k-1}$ of line 12 of  Algorithm \ref{alg:CG}, which yields
\begin{align} \label{eq:rk}
\bmr_{k}=\bmr_{k-1}-\alpha_{k}\bA\bmp_{k-1}. 
\end{align}
By rewriting line 14 of Algorithm \ref{alg:CG}  as  $\bmr_{k} = \bmp_{k} - \beta_k\bmp_{k-1}$, we can substitute $\bmr_{k}$ by $\bmp_{k} - \beta_k\bmp_{k-1}$ in  (\ref{eq:rk}), which leads to
\begin{align}\label{eq:pk}
\bmp_{k}=\bmp_{k-1}+\beta_{k}\bmp_{k-1}-\alpha_{k}\bA\bmp_{k-1}-\beta_{k-1}\bmp_{k-2}. 
\end{align}
By rewriting line 11 of  Algorithm \ref{alg:CG} to  $\bmp_{k-1} = (\bmv_{k}- \bmv_{k-1})/\alpha_k$, we can replace $\bmp_{k-1}$ with $(\bmv_{k}- \bmv_{k-1})/\alpha_k$ in  (\ref{eq:pk}). Then, by using $\bmv_k=\bmxhat_k$ of line 19, we can formulate the desired recursion (for $k=1,\ldots,K$):
\begin{align*}
\bmxhat_{k}=\,\,&  \bmxhat_{k-1}+\left(\frac{\alpha_{k}(1+\beta_{k-1})}{\alpha_{k-1}}\bI_U-\alpha_{k}\bA\right)(\bmxhat_{k-1} - \bmxhat_{k-2})  \\
&-\frac{\alpha_{k}\beta_{k-2}}{\alpha_{k-2}}(\bmxhat_{k-2} - \bmxhat_{k-3}).
\end{align*}
Here, $\bmxhat_1=\alpha_1\bH_u^H\bmy$, and $\alpha_k$ and $\beta_k$ for $k=1,\ldots,K$ are computed in Algorithm~\ref{alg:CG}; in addition, we initialize $\alpha_{k}=1$, $\beta_{k}=\,0$, and $\bmxhat_k=\bZero_{\MT\times1}$ for $k < 1$.

Since  $\bmxhat_k$ can be computed from $\bL_{k}\bH^{H}_u\bmy$ as in \eqref{eq:eqmatrix}, the   matrix~$\bL_{k}$ can be obtained recursively  as follows:
\begin{align}
\bL_{k}=&\,\bL_{k-1}+\left(\frac{\alpha_{k}(1+\beta_{k-1})}{\alpha_{k-1}}\bI_U-\alpha_{k}\bA\right)(\bL_{k-1} - \bL_{k-2}) \notag \\
&-\frac{\alpha_{k}\beta_{k-2}}{\alpha_{k-2}}(\bL_{k-2} - \bL_{k-3}), \label{eq:Lmatrix}
\end{align}
where we set $\bL_{1}=\alpha_1\bI_U$ and $\bL_{k}=\bZero_{\MT\times\MT}$ for $k < 1$. 
This recursion allows us to compute $\mu_{i|k}$  and $\nu_{i|k}^2$ in each iteration~$k$ of the CG procedure in Algorithm~\ref{alg:CG}. 

Specifically, let $\bB= \bL_{k}\bG$ with the Gram matrix $\bG=\bH^H_u\bH_u$, then $\mu_{i|k}$ can be computed as $\mu_{i|k} = B_{i,i}$, where $ B_{ii}$ is the $i^\text{th}$ diagonal entry of $\bB$.
To compute $\nu_{i|k}^2$, let $\bC = \bB\bL_k^H$ and compute the variance of interference plus noise as
\begin{align*}
\nu_{i|k}^2 = \textstyle \sum_{j,j\neq i}|B_{i,j}|^2\Es + C_{i,i} \No,
\end{align*}
where $C_{i,i}$ is the $i^\text{th}$ diagonal entry of $\bC$. We can now compute LLR values in \eqref{eq:llr} using the quantities $\mu_{i|k}$ and $\rho_{i|k} = \mu_{i|k}^2/{\nu_{i|k}^2}$ obtained in each CG iteration $k$. Note that if $k=\MT$, then this SINR tracking scheme provides exact results, i.e., we have $\mu_{i}=\mu_{i|\MT}$ and $\rho_{i}=\rho_{i|\MT}$, and Algorithm \ref{alg:CG} can be used to compute \eqref{eq:llr} exactly. When terminating the CG procedure early, we can still approximate the LLR values in \eqref{eq:llr} but at (often significantly) lower computational complexity  (see our simulation results in Section~\ref{sec:sims}).

\subsubsection{Approximate SINR Computation}

The above exact SINR computation method \eqref{eq:Lmatrix} requires a $\MT\times\MT$ matrix multiplication per iteration, which adds high computational complexity to the (otherwise low-complexity) CG method.
To reduce the overall computational complexity, we next propose an approximate method that is very accurate for massive MIMO systems.

We start by noting that the regularized Gram matrix $\bA = \bH^H_u\bH_u+\varrho_u^{-1}\bI_U$ is diagonally dominant for massive MIMO systems~\cite{Wu2012,Yin2013,Wu2014}. Hence, its main diagonal $\bD$ well approximates~$\bA$. We now exploit this property to approximate~$\bL_{k}$ in   \eqref{eq:Lmatrix} by replacing $\bA$ with $\bD$ to obtain the the following approximate recursion (for $k=1,\ldots,K$):
\begin{align} 
\widetilde{\bL}_{k} = & \,\, \widetilde{\bL}_{k-1}+\left(\frac{\alpha_{k}(1+\beta_{k-1})}{\alpha_{k-1}}\bI_U-\alpha_{k}\bD\right)(\widetilde{\bL}_{k-1} - \widetilde{\bL}_{k-2}) \notag \\
&-\frac{\alpha_{k}\beta_{k-2}}{\alpha_{k-2}}(\widetilde{\bL}_{k-2} - \widetilde{\bL}_{k-3}), \label{eq:lowcomplexitytracker}
\end{align}
which we initialize with $\widetilde{\bL}_{1}=\alpha_1\bI_U$ and $\widetilde{\bL}_{k}=\bZero_{\MT\times\MT}$ for $k < 1$. 
Since all matrices in the recursion \eqref{eq:lowcomplexitytracker} are (and remain to be) diagonal,  this approximate method  only requires $\MT$ multiplications per CG iteration.

From the approximate matrix $\widetilde{\bL}_{k}$ in \eqref{eq:lowcomplexitytracker}, we can compute an approximate of the quantity $\mu_{i|k}$ in iteration $k$ as
\begin{align}  \label{eq:mu}
\mu_{i|k} \approx \widetilde{L}_{i,i|k}G_{i,i},
\end{align}
where $G_{i,i}$ is the $i^\text{th}$ diagonal entry of  $\bG$, and  $ \widetilde{L}_{i,i|k}$ is the $i^\text{th}$ diagonal entry of the    matrix $\widetilde{\bL}_k$.
An approximate for $\nu_{i|k}^2$ is obtained analogously by computing
$\nu_{i|k}^2\approx \No \widetilde{L}_{i,i|k}^2G_{i,i}$,
which we use to approximate the post-equalization SINR as 
\begin{align}\label{eq:rho}
\rho_{i|k}={\mu_{i|k}^2}/{\nu_{i|k}^2}\approx {G_{i,i}}/ {\No},
\end{align}
which complexity does not depend on the iteration index $k$.

\subsection{CG-P: CG-Based Linear Precoding}
\label{sec:cgprecoder}
Similar to the uplink, CG can be used for downlink precoding, i.e., to compute \eqref{eq:precoding}.
Unlike soft-output detection, precoding via CG is rather straightforward. In particular, we solve
\begin{align} \label{eq:CGdownlink}
\bmv = \underset{\bmvtilde\in\mathbb{C}^\MT}{\operatorname{arg\,min}} \,  \|\bmt - \bA\bmvtilde\|_2, 
\end{align}
with $\bA= \bH_d\bH_d^H+\varrho^{-1}_d\bI_\MT$ and then, compute the precoded vector as $\bmq = {\bH^H_d}\bmv$.
The precoding problem \eqref{eq:CGdownlink} can be solved efficiently using Algorithm \ref{alg:CG}. 
Since precoding does not require the post-equalization SINR, it requires lower complexity than CG for soft-output detection.

Since CG is an exact method, our algorithm performs exact MMSE precoding \eqref{eq:precoding} at iteration $K=\MT$.
For $K=1$, CG-based precoding corresponds to matched filter precoding $\bmq=\alpha_1{\mathbf{H}}^H\mathbf{t}$ with $\alpha_1$ computed in line 10 of Algorithm \ref{alg:CG}.
In the large-antenna limit, matched-filter precoding is known to be optimal~\cite{Marzetta2010}. Hence, in massive MIMO systems, only few CG iterations are sufficient to achieve near-optimal performance; corresponding results are shown in Section \ref{sec:sims}. 

\subsection{What About CG Least Squares?}

Another well-known variant of CG is the so-called conjugate gradient least squares~(CGLS) method~\cite{Paige1982}.
In contrast to regular CG, this variant is capable of operating with non-square matrices, which avoids computation of the regularized Gram matrix. 
Consequently,  CGLS can be used for uplink detection by solving 
\begin{align} \label{eq:CGLSuplink}
\bmxhat = \underset{\bmxtilde\in\mathbb{C}^{\MT}}{\operatorname{arg\,min}} \,  \|\overline{\bmy} - \overline{\bH}_u\bmxtilde\|, 
\end{align}
where $\overline{\bmy}=\big[ \bmy^T , \bZero_{1\times\MT}\big]^T$ and $\overline{\bH}_u=\big[\bH^T_u, \sqrt{\varrho_u^{-1}}\bI_\MT\big]^T$ are the augmented receive vector and uplink channel matrix, respectively. Note that the recursive SINR computation methods of Section~\ref{sec:SINRtrackers} can be built into CGLS as well. 
CGLS can also be used for precoding by solving
\begin{align} \label{eq:CGLSdownlink}
\overline{\bmq} = \underset{\bmqtilde\in\mathbb{C}^{\MR+\MT}}{\operatorname{arg\,min}} \,  \|\bmt - \overline{\bH}_d\bmqtilde\|, 
\end{align}
where $\overline{\bH}_d=\big[ \bH_d , \sqrt{\varrho_d^{-1}}\bI_\MT\big]$ is the augmented downlink channel matrix. The  precoded vector can then be extracted from~$\overline{\bmq}$ as follows $\bmq = [\overline{q}_1,\ldots,\overline{q}_B]^T$.
We note that since CGLS avoids computation of the Gram matrix, it has the potential to require even lower complexity than CG. 
Unfortunately, as we will show in Section~\ref{sec:sims}, CGLS is only advantageous over CG for a very small number of iterations, where both methods typically deliver sub-optimal error-rate performance.

\subsection{Other Low-Complexity Detection and Precoding Methods}

We next summarize other exact and approximate inversion methods, which we use in Section~\ref{sec:sims} to compare to our CG-based soft-output detection and precoding method. 

\subsubsection{Exact Matrix Inversion}
The Cholesky decomposition~\cite{schreiber1986systolic} is among the  computationally efficient ways of exactly computing the matrix inverse required for uplink detection \eqref{eq:eqweight} and downlink precoding  \eqref{eq:precodingmatrix} (see, e.g., \cite{Wu2014}  for a reference design).  
A Cholesky-based matrix inversion proceeds as follows. 
First, one computes the regularized Gram matrix $\bA_u = {\bH}_u^H{\bH}_u+\varrho_{u}^{-1}\bI_\MT$ or $\bA_d = {\bH}_d{\bH}_d^H+\varrho_{d}^{-1}\bI_\MT$ for the uplink or downlink case, respectively. 
Then, the regularized Gram matrix is decomposed as $\bA=\bM\bM^H$, where $\bM$ is lower-triangular (we omit the subscripts $_d$ and $_u$). Then, one can perform a forward- and backward substitution procedure to obtain $\bA^{-1}=(\bM\bM^H)^{-1}$ in a computationally efficient manner (see \cite{Studer2011} for the algorithm details). 

\subsubsection{Approximate Matrix Inversion}

To reduce the computational complexity of an exact matrix inversion (for detection or precoding), approximate inversion schemes have been investigated in \cite{Wu2014,Wu2012,Yin2013,YWWDCS14b}. 
The main idea is to use a truncated Neumann series expansion, which proceeds as follows. 
First, one computes the regularized Gram matrix $\bA_u$ or $\bA_d$, which is then decomposed into a diagonal and off-diagonal part according to $\bA = \bD+\bE$. 
A truncated $K$-term Neumann series expansion is then used to obtain an approximate inverse~\cite{Wu2014}:
\begin{align*}
\bA^{-1} \approx \textstyle \sum_{k=0}^{K-1}({-\bD}^{-1}\bE)^k{\bD}^{-1}.
\end{align*}
For a small number of Neumann series terms, i.e., for $K\leq 3$, this approximation requires very low complexity.

\section{Simulation Results}
\label{sec:sims}
 
  \begin{figure*}[tp]
\centering
\subfigure[$\MR=32$ and $\MT=8$]{\includegraphics[height=0.41\columnwidth]{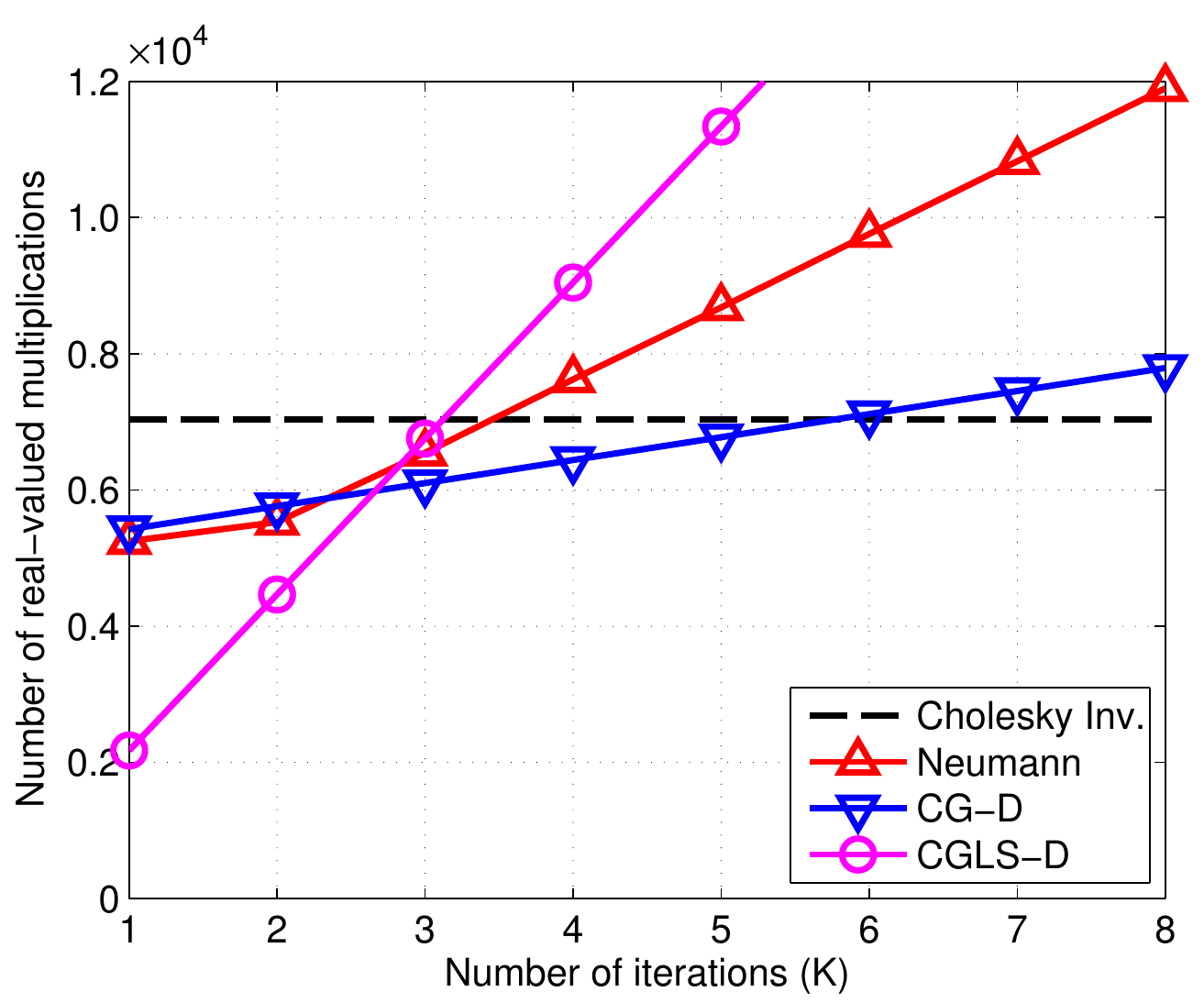}\label{fig:32x8_Complexity}}
\subfigure[$\MR=128$ and $\MT=8$]{\includegraphics[height=0.41\columnwidth]{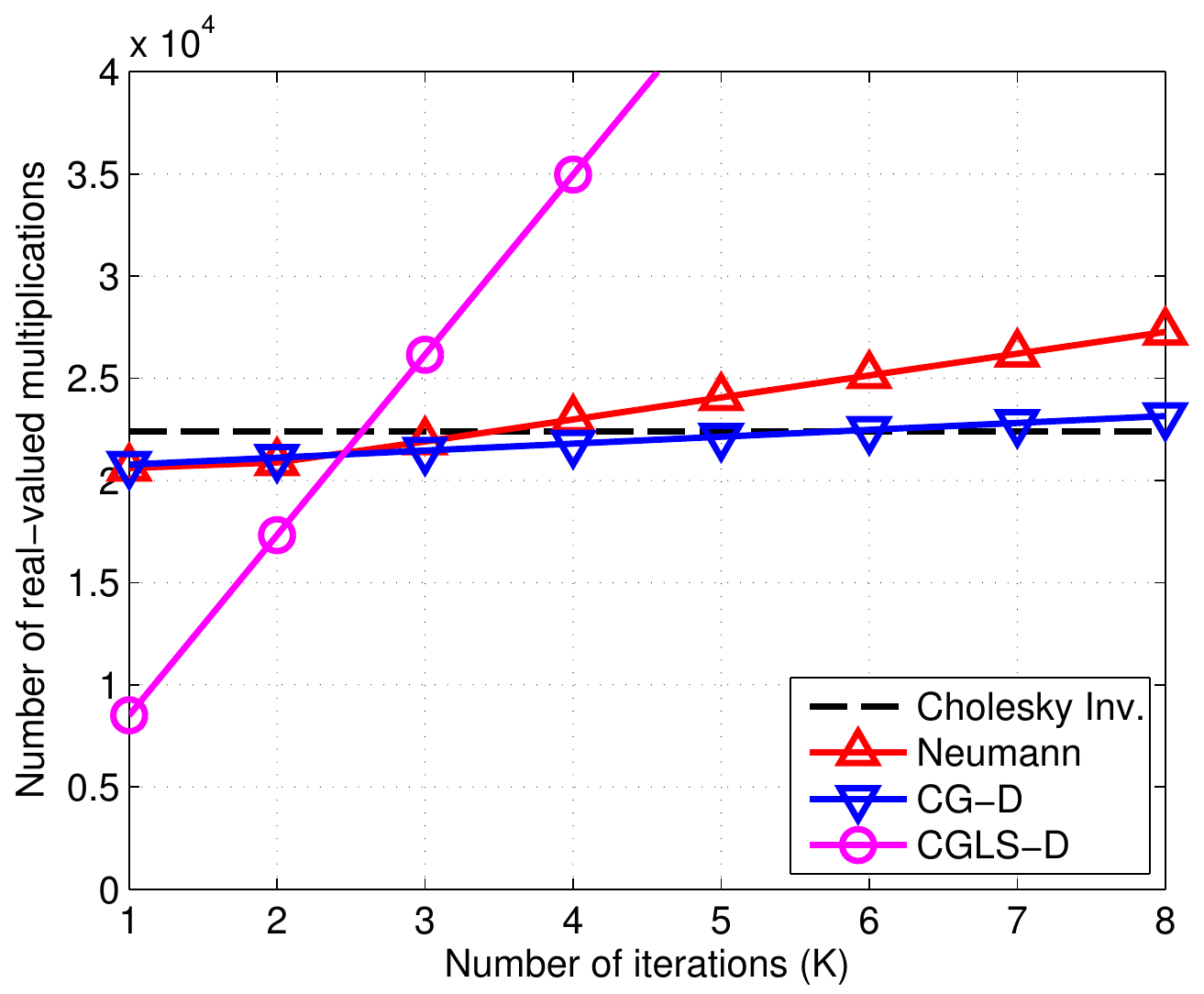}\label{fig:128x8_Complexity}}
\subfigure[$\MR=32$ and $\MT=16$]{\includegraphics[height=0.41\columnwidth]{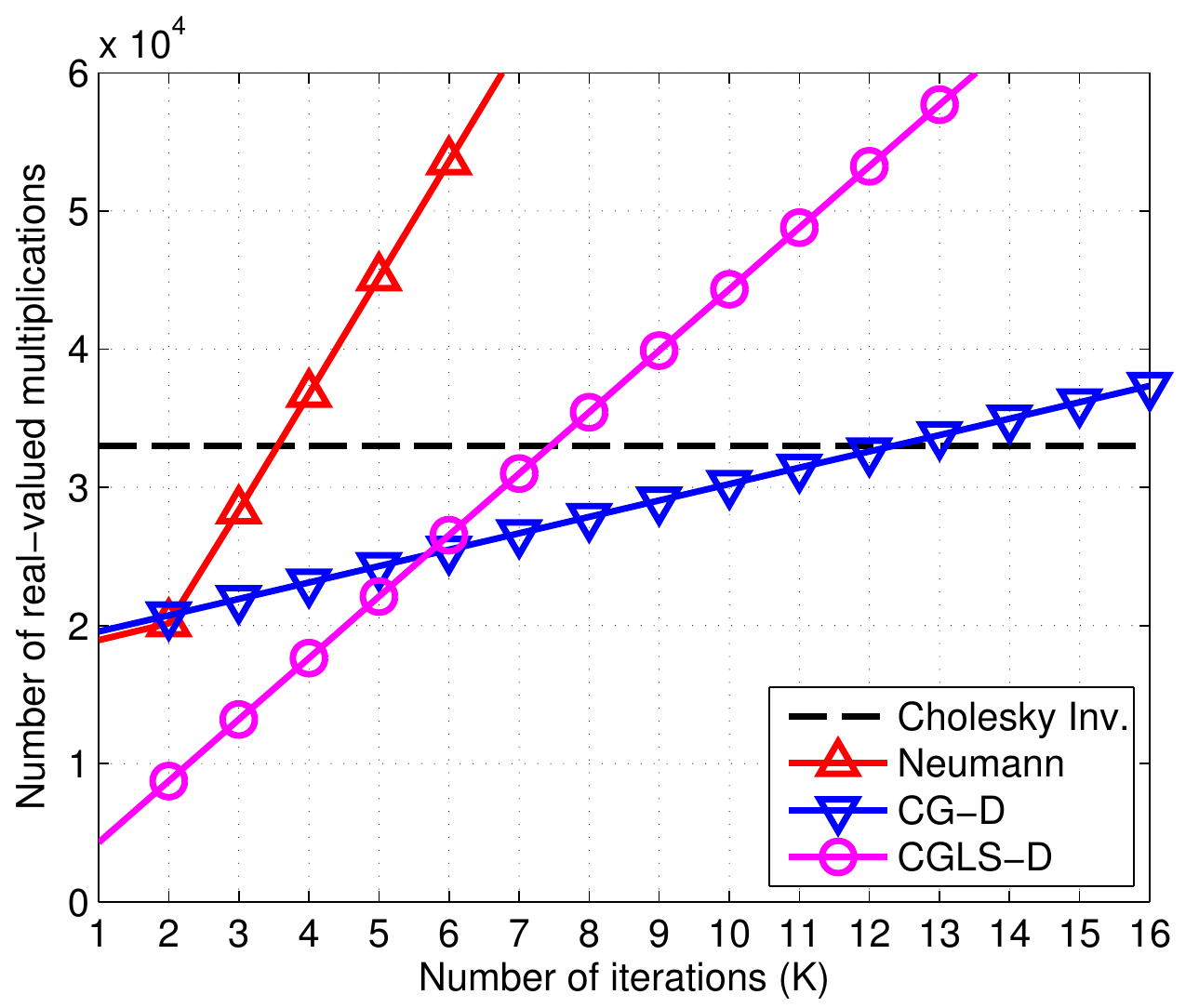}\label{fig:32x16_Complexity}}
\subfigure[$\MR=128$ and $\MT=16$]{\includegraphics[height=0.41\columnwidth]{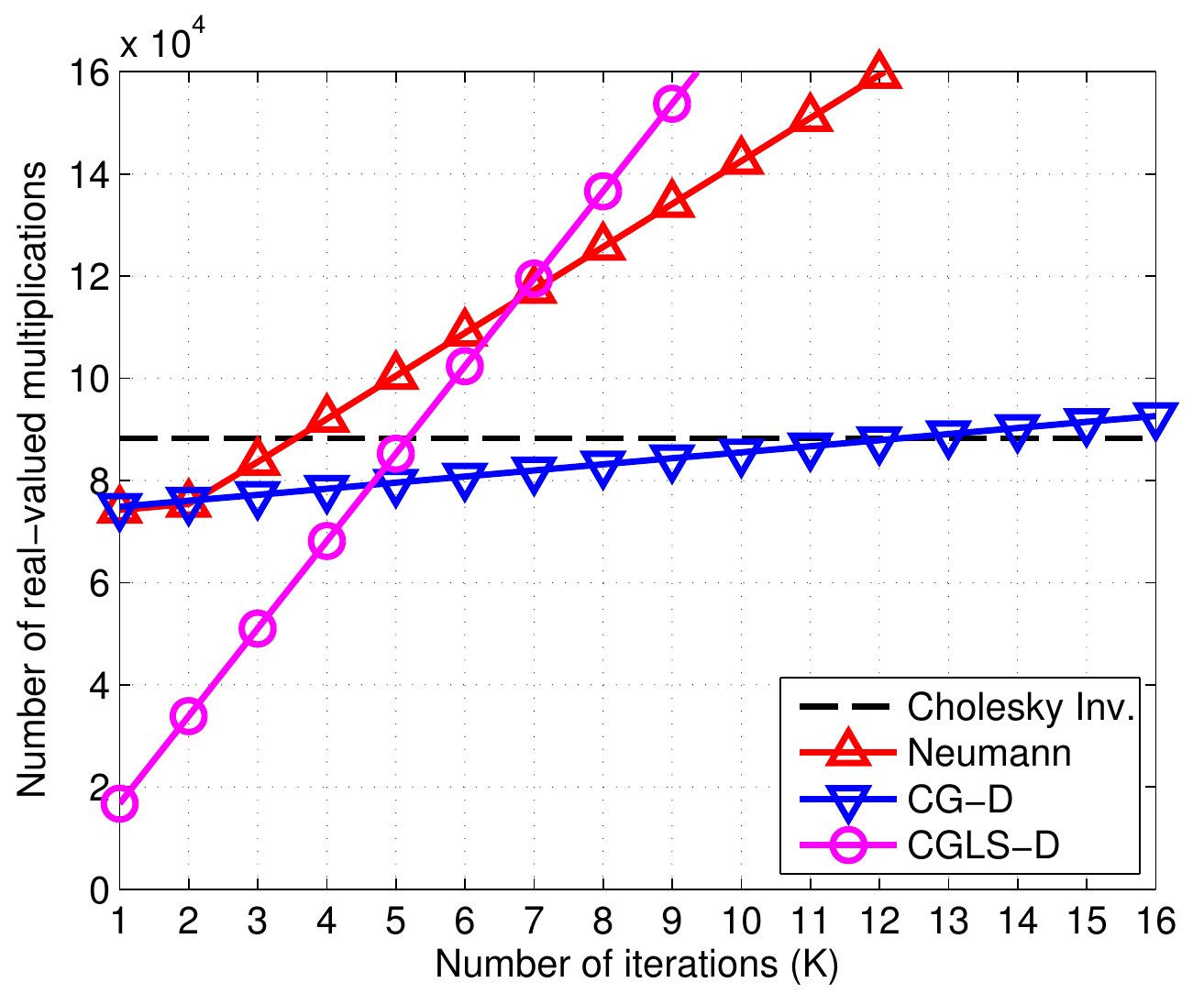}\label{fig:128x16_Complexity}}
\vspace{-0.05 in}
\caption{Uplink detection complexity comparison between the reference Cholesky-based soft-output MMSE detector, the Neumann series approach in \cite{Wu2014}, and the proposed CG and CGLS-based soft-output detectors (CG-D). We measure the complexity in terms of the number of real-valued multiplications.}
\label{fig:complexity}
\vspace{-0.1 in}
\end{figure*}

\begin{figure*}[tp]
\centering
\subfigure[$\MR=32$ and $\MT=8$]{\includegraphics[width=0.5\columnwidth]{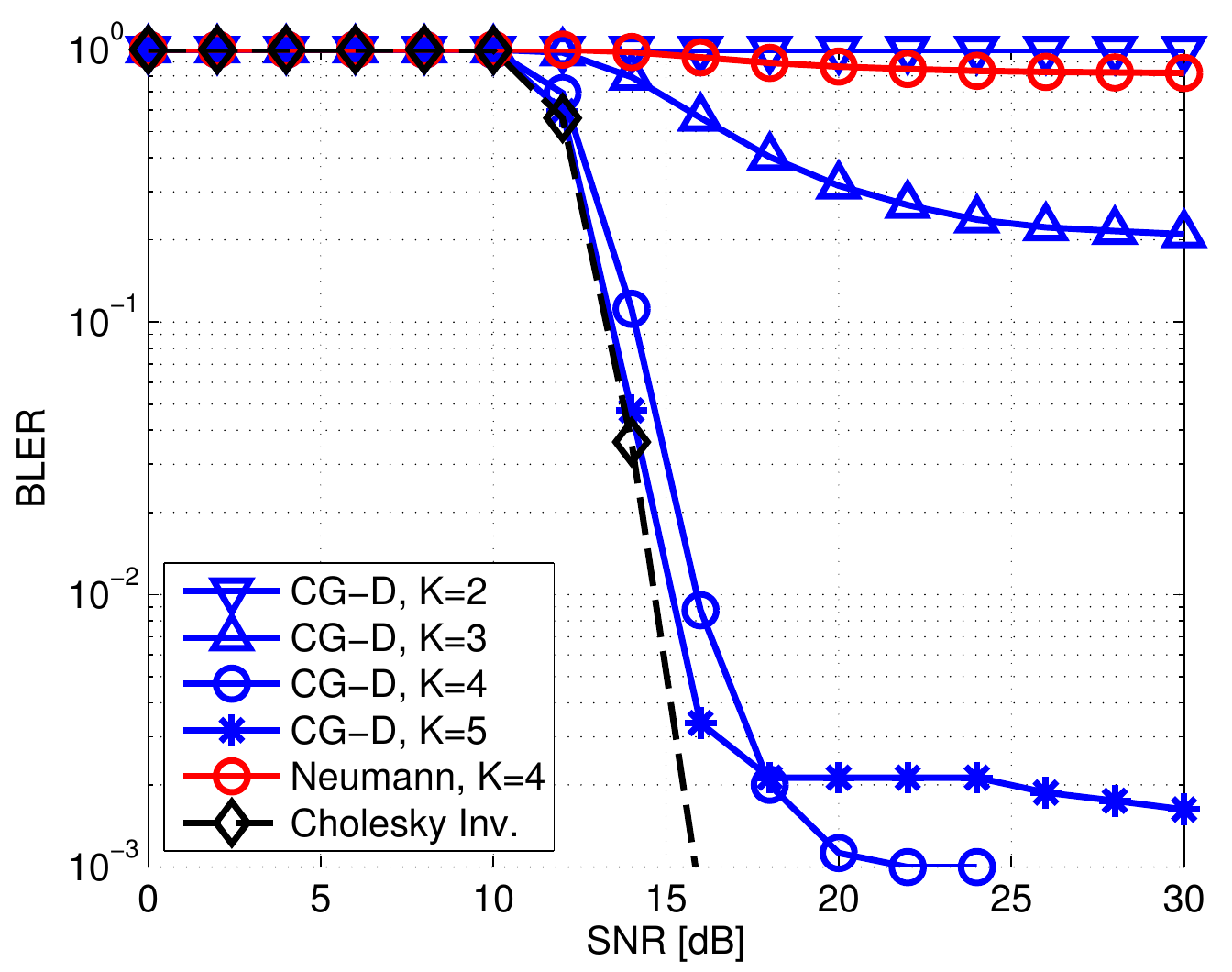}\label{fig:bler_bcjr_aproxmu_aproxvar_32_8_64qam}}
\subfigure[$\MR=128$ and $\MT=8$]{\includegraphics[width=0.5\columnwidth]{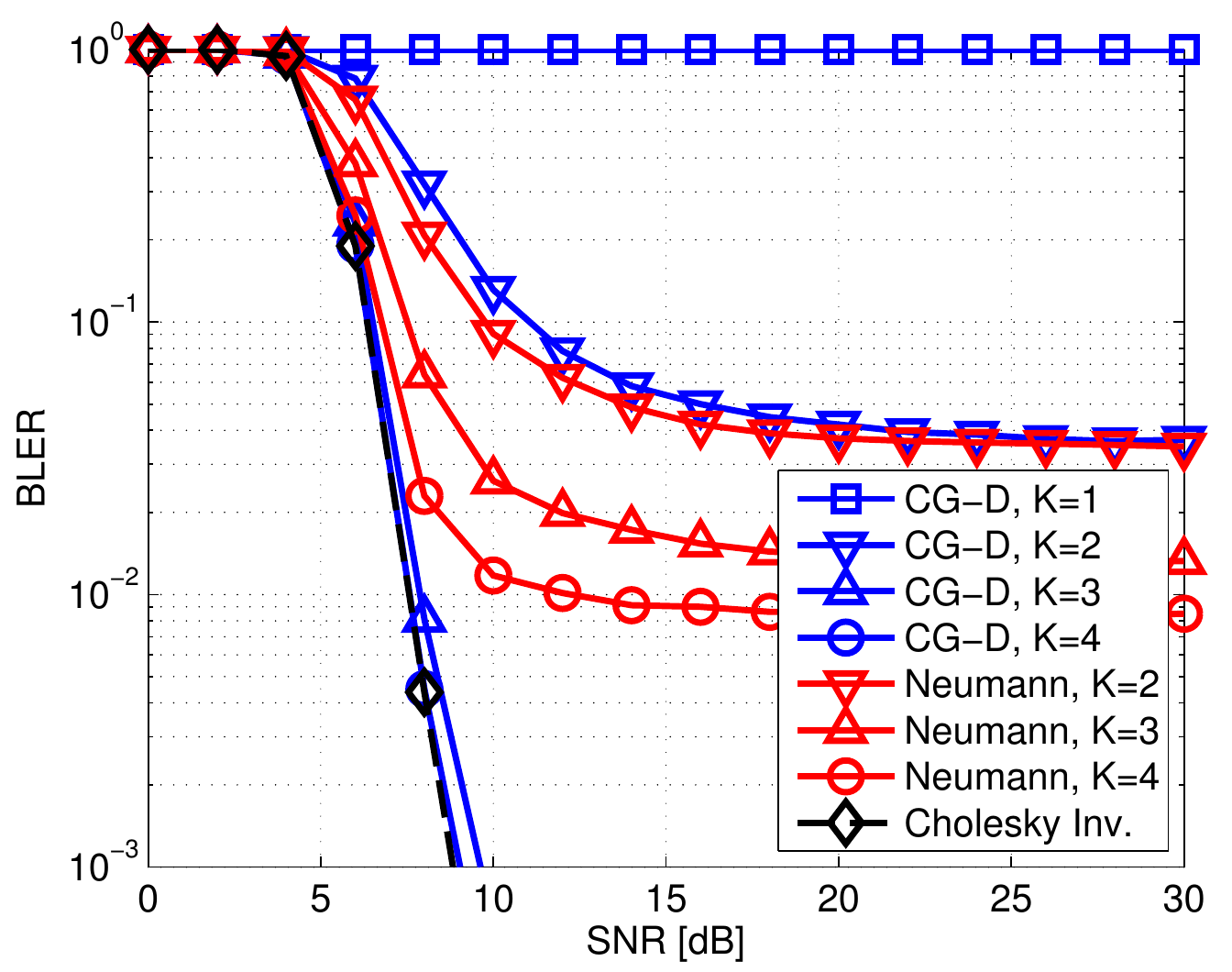}\label{fig:bler_bcjr_aproxmu_aproxvar_128_8_64qam}}
\subfigure[$\MR=32$ and $\MT=16$]{\includegraphics[width=0.5\columnwidth]{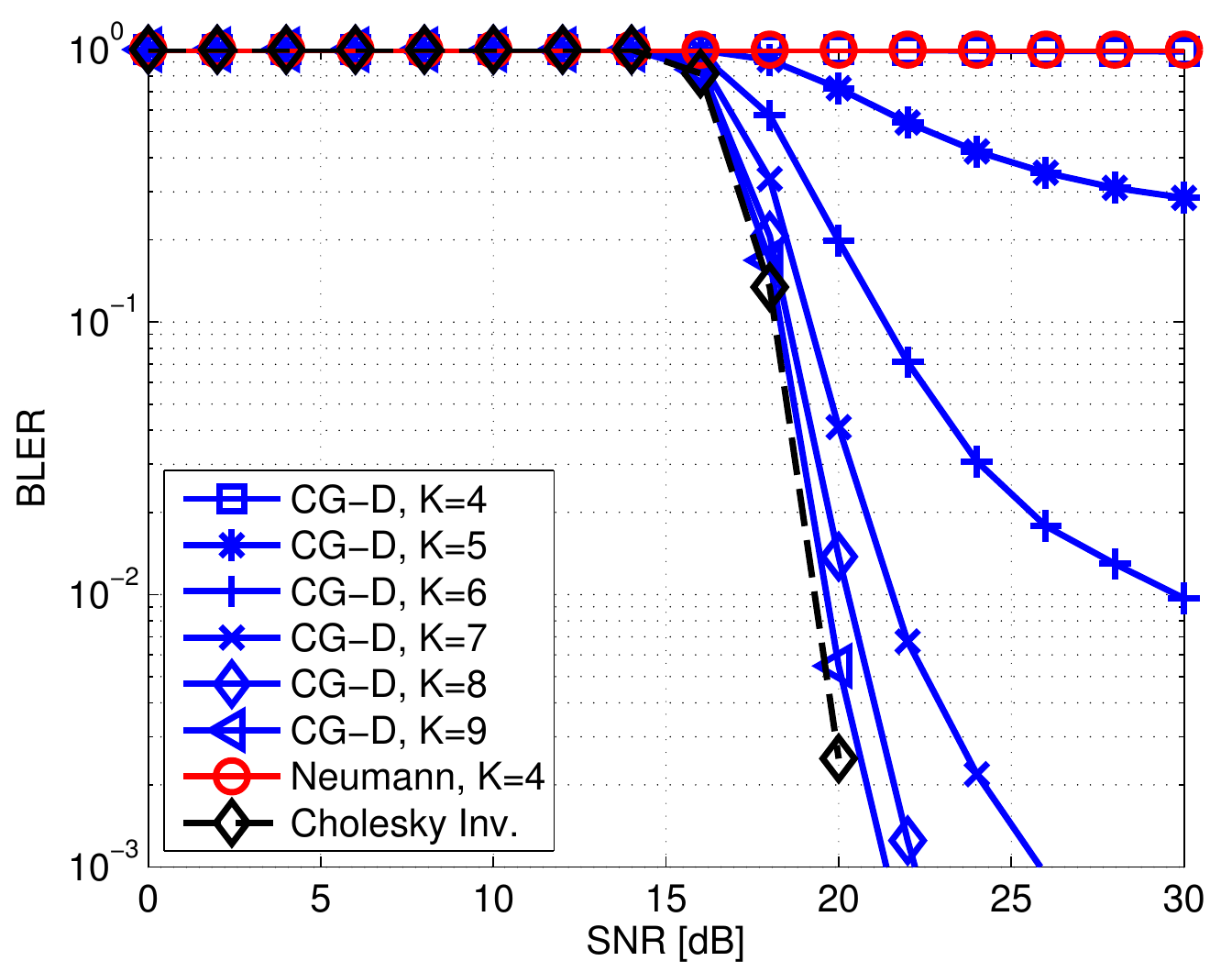}\label{fig:bler_bcjr_aproxmu_aproxvar_32_16_64qam}}
\subfigure[$\MR=128$ and $\MT=16$]{\includegraphics[width=0.5\columnwidth]{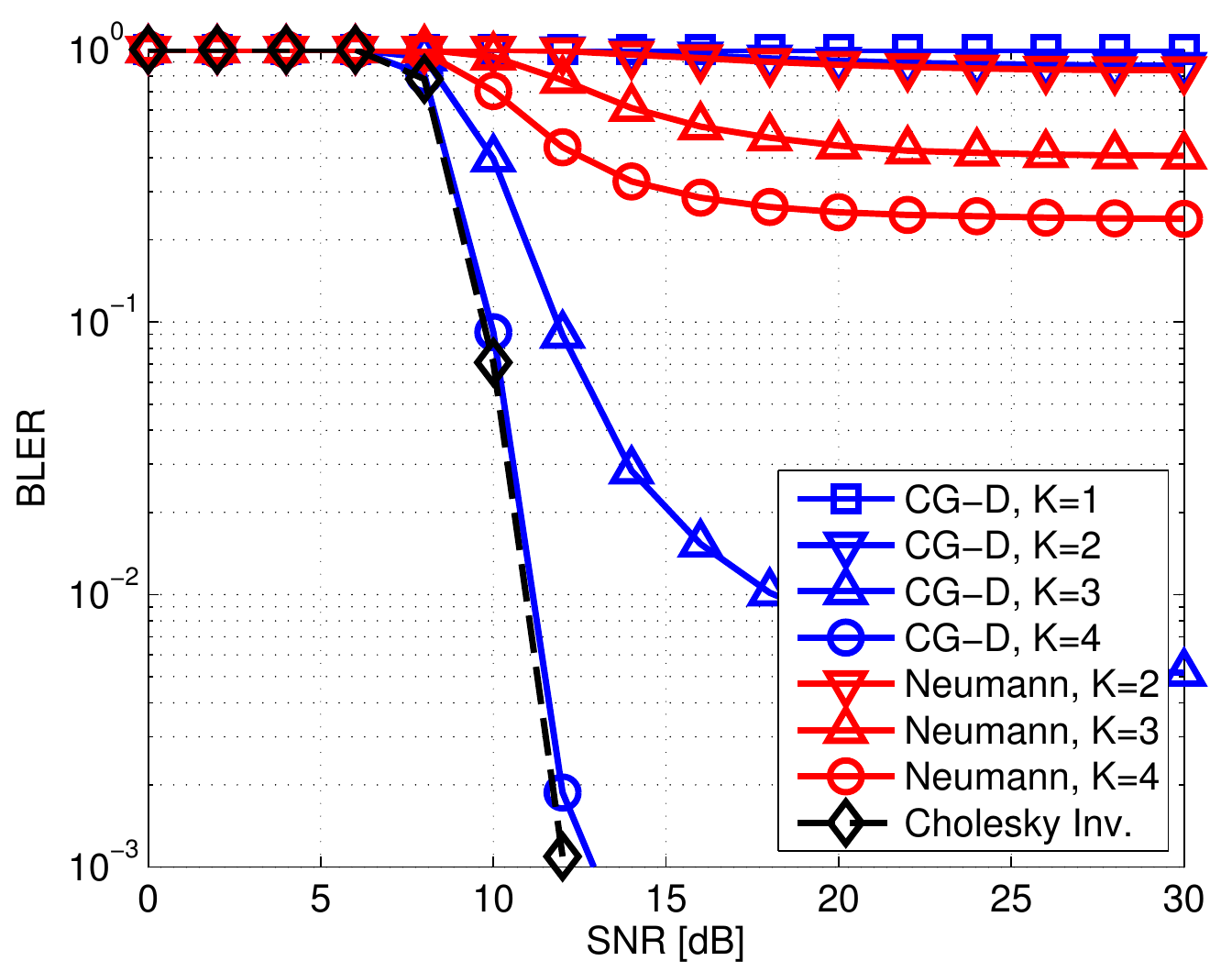}\label{fig:bler_bcjr_aproxmu_aproxvar_128_16_64qam}}
\caption{Block error-rate (BLER) performance comparison in the massive MIMO uplink for the reference Cholesky-based soft-output MMSE detector, the Neumann series approach in \cite{Wu2014}, and the proposed CG-based soft-output detector (CG-D). We note that CGLS achieves the same BLER as CG.}
\label{fig:BLER_ul}
\end{figure*}

We now showcase the efficacy and limits of the proposed CG-based soft-output detection and precoding approach. In the ensuing discussion, we solely consider the approximate SINR computation strategy proposed in Section~\ref{sec:SINRtrackers}.

\subsection{Computational Complexity Analysis}
\label{sec:complexity}
The computational complexity of the exact Cholesky-based matrix inversion, the approximate Neumann series approach~\cite{Wu2014}, and our CG and CGLS-based methods is dominated by multiplications. Hence, we compare the complexity of these methods by counting the number of real-valued\footnote{We count $4$ real-valued multiplications per complex-valued multiplication.}  multiplications.
For all algorithms, we exploit properties of Hermitian matrices and avoid multiplications with zeros.
 
Figures~\ref{fig:complexity}(a)--(d) compare the computational complexity for massive MIMO systems with the following  BS $\times$ user antenna configurations: $32\times8$, $128\times8$, $32\times16$, and $128\times16$.
For the $\MT=8$ user configurations, CG-based soft-output detection requires lower complexity than the exact Cholesky-based inversion when $k\leq5$ (Figs.~\ref{fig:32x8_Complexity} and \ref{fig:128x8_Complexity}). 
Similarly,  for the $\MT=16$ user configurations, CG-based detection requires lower complexity than the exact Cholesky-based approach when $k\leq12$  (Figs.~\ref{fig:32x16_Complexity} and \ref{fig:128x16_Complexity}).
Although CGLS and the Neumann series approximation exhibit lower complexity than our CG-based algorithm for a small number of iterations~$K$, we next show that the error-rate performance for these two methods is inferior to that of CG-based detection. 

We note that CG-based methods as in Algorithm \ref{alg:CG} and the Neumann approach exhibit higher regularity than the Cholesky-based approach, which will enable more efficient hardware designs (see also \cite{Wu2014}); this distinct advantage is not reflected in the simplistic complexity measure considered here. 

\subsection{Block Error-Rate (BLER) Performance}
\label{sec:BLERcomparison}

\begin{figure*}[tp]
\centering
\subfigure[$16$-QAM and $\MT=8$]{\includegraphics[height=0.5\columnwidth]{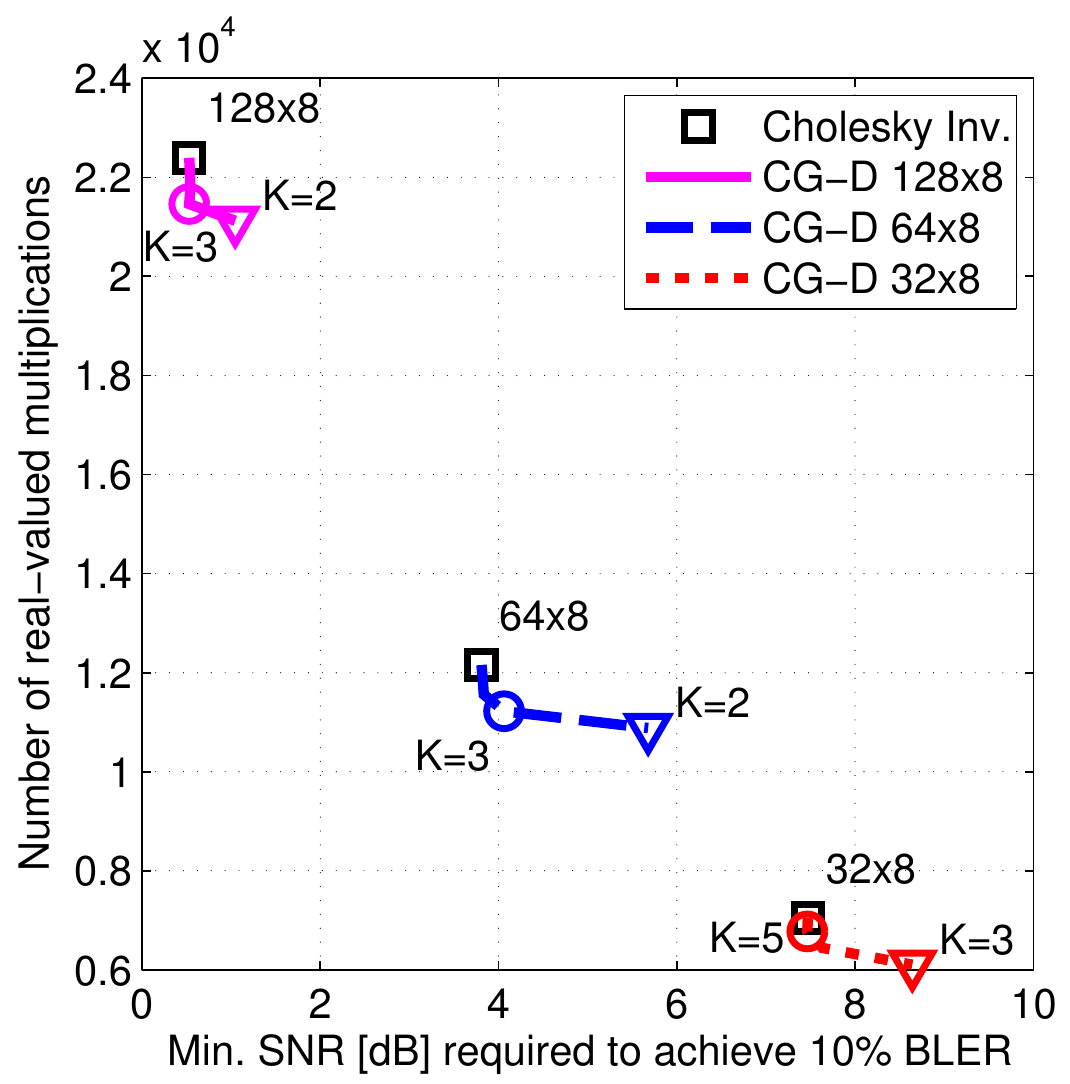}\label{fig:tradeoff16_8u}}
\subfigure[$16$-QAM and $\MT=16$]{\includegraphics[height=0.5\columnwidth]{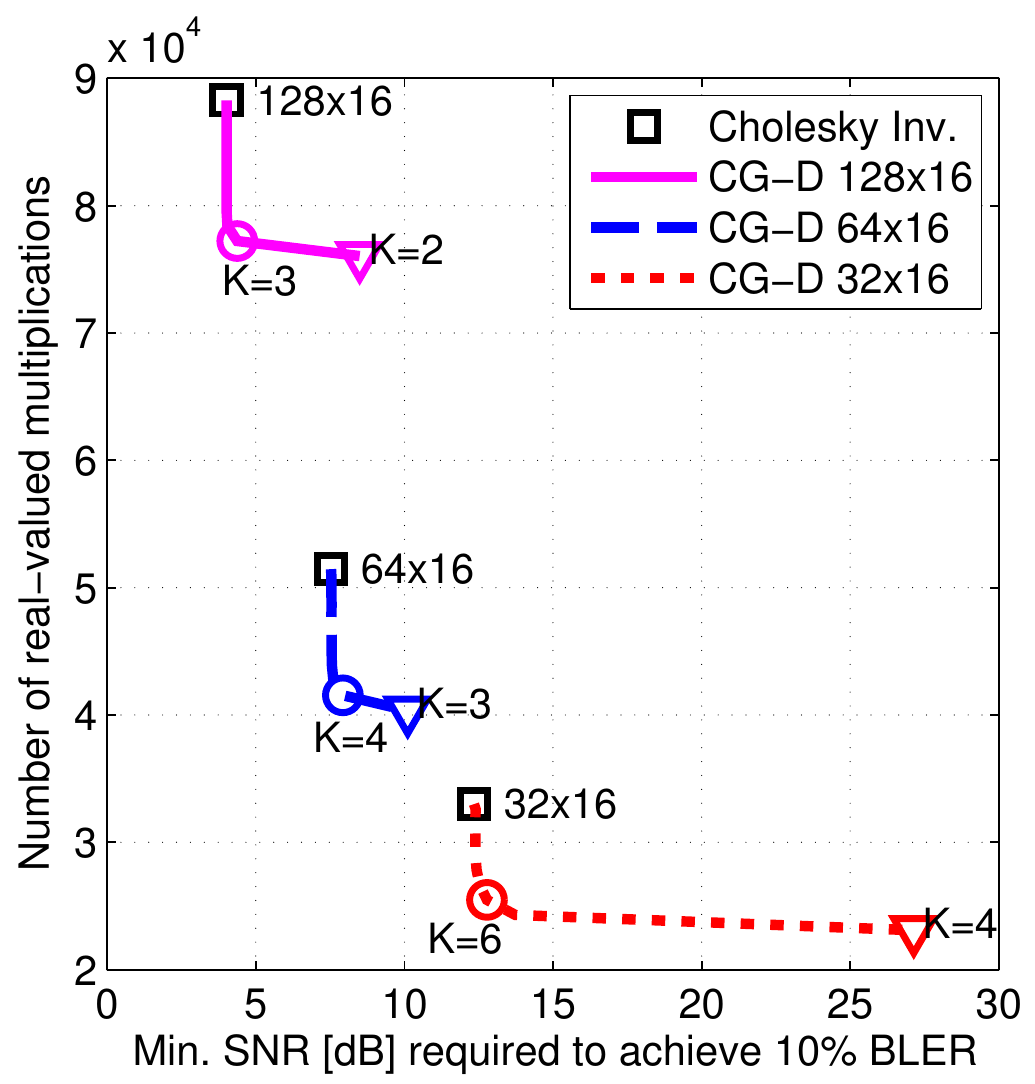}\label{fig:tradeoff16_16u}}
\subfigure[$64$-QAM and $\MT=8$]{\includegraphics[height=0.5\columnwidth]{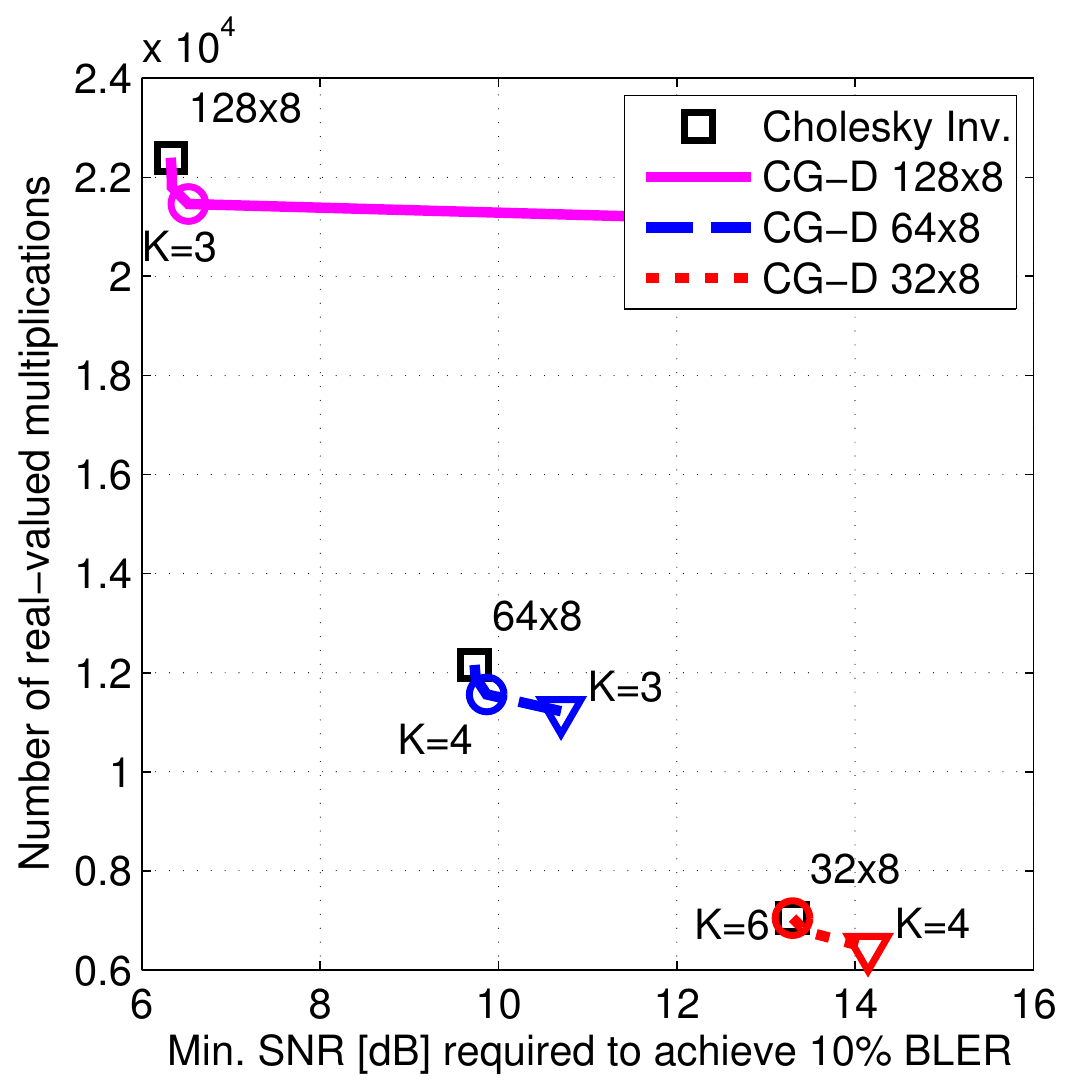}\label{fig:tradeoff64_8u}}
\subfigure[$64$-QAM and $\MT=16$]{\includegraphics[height=0.5\columnwidth]{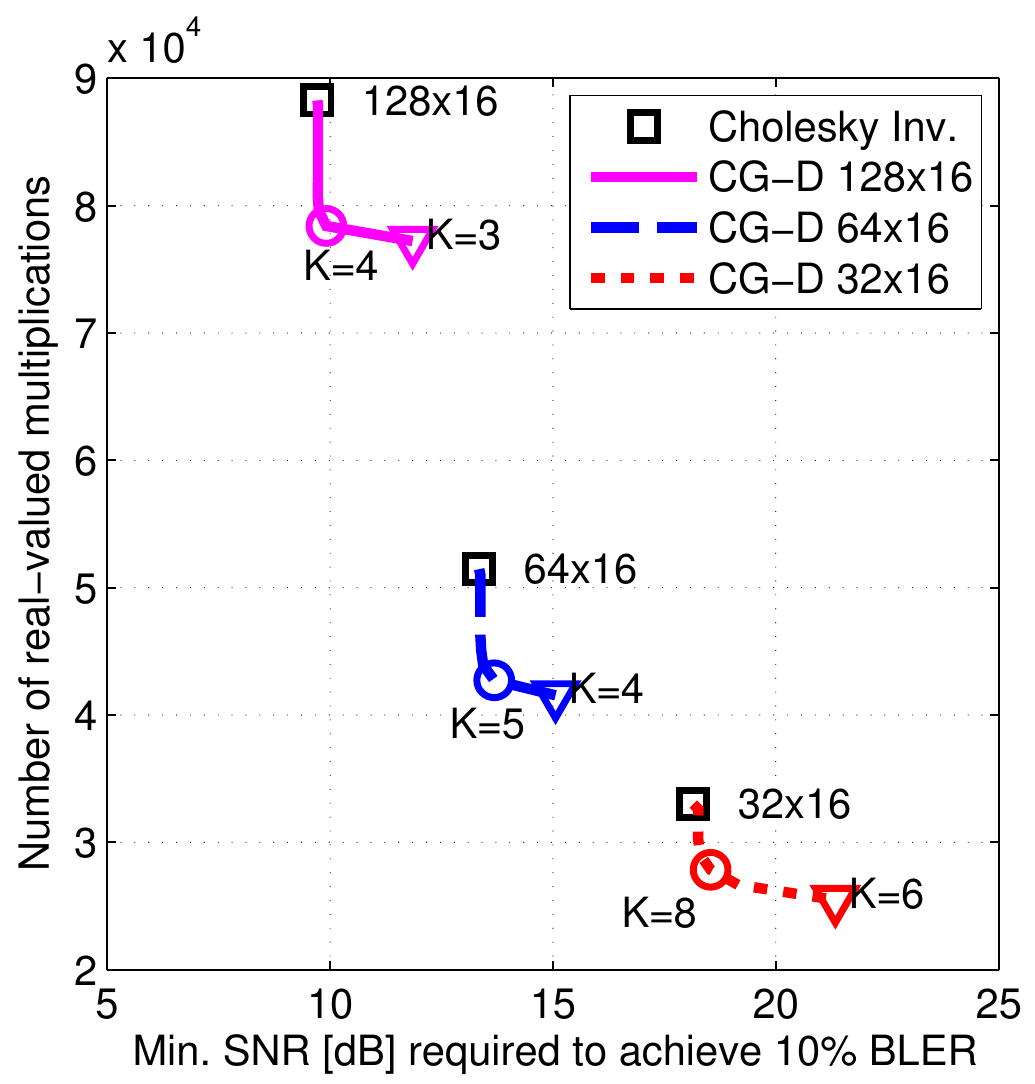}\label{fig:tradeoff64_16u}}
\caption{Performance/complexity trade-off of the proposed CG-based soft-output detector (CG-D) and an exact Cholesky-based soft-output detector.}
\label{fig:tradeoff}
\end{figure*}

To compare the block-error rate (BLER) performance of the proposed algorithms with existing methods, we simulate an OFDM system with 128 subcarriers and a ${5/6}$-rate convolutional code. At the BS, we deploy a $10$\,m linear antenna array with equally-spaced antennas. The corresponding channel matrices are generated using WINNER-Phase-2 model~\cite{winner2}.\footnote{At the time, we are unaware of any channel model for massive MIMO.}
For the uplink, the BS performs soft-output MMSE detection; for the downlink we perform linear MMSE precoding. 
In both cases, we deploy a soft-input max-log Viterbi decoder either at the BS (for the uplink) or at each user (for the downlink). 

The resulting uplink BLER performance is shown in Figs.~\ref{fig:BLER_ul}(a)--(d). We see that our CG-based soft-output detector achieves a BLER that is close to the (exact) reference Cholesky method $K=5$, $K=3$, $K=8$, and $K=4$  for the ${32\times 8}$,  ${128\times 8}$,  ${32\times 16}$, and ${128\times 16}$ antenna configuration. In all these cases, the associated computational complexity is lower to that of the exact inversion method. 
Note that we omit the BLER performance of the CGLS method, as CGLS and CG deliver the same outputs, which leads  to the same performance.
In addition, we see that the Neumann series approach exhibits a rather high error floor for the considered antenna configurations. The Neumann series works well in the $128\times8$ case, which confirms the observation in \cite{Wu2014} that this approach requires systems having a large BS to user antenna ratio.

\subsection{Performance/Complexity Trade-offs}
\label{sec:tradeoffs}

The overall advantage of CG-based soft-output detection is summarized in Figs.~\ref{fig:tradeoff}(a)--(d), where we plot the trade-off between complexity and performance, measured in terms of the minimum SNR required to achieve $10$\% BLER. 
The Neumann series approach is omitted due to the rather high error floor.
We see that CG-based detection is able to achieve lower complexity and similar error-rate performance as the exact Cholesky-based detection for all considered cases. Note that the complexity savings (for an equal SNR performance) are more pronounced for the $\MT=16$ user antenna cases. 

\begin{figure}[tp] 
\centering
\subfigure[$64$-QAM and $\MT=8$]{\includegraphics[height=0.5\columnwidth]{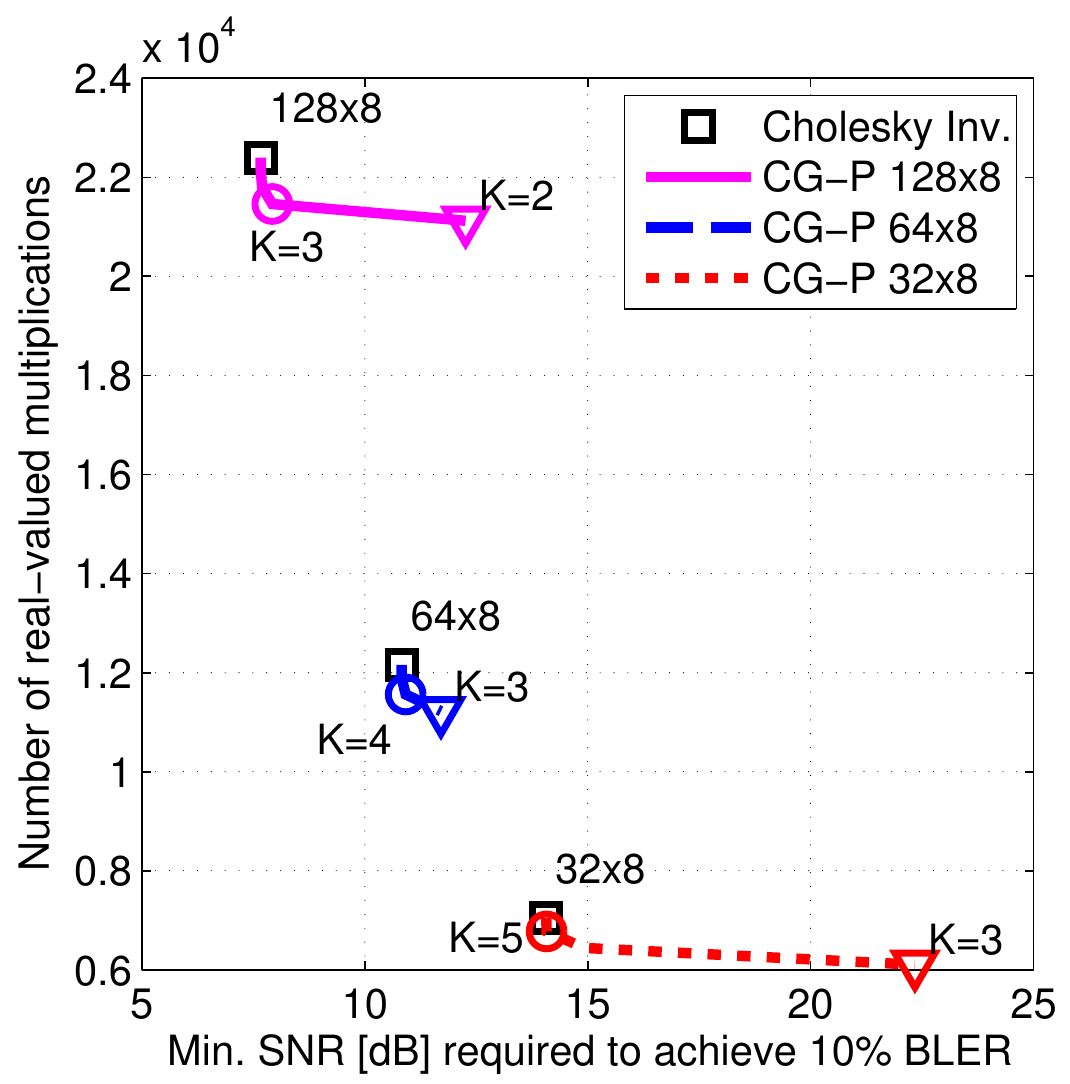}\label{fig:tradeoff64_8u_dl}}
\subfigure[$64$-QAM and $\MT=16$]{\includegraphics[height=0.5\columnwidth]{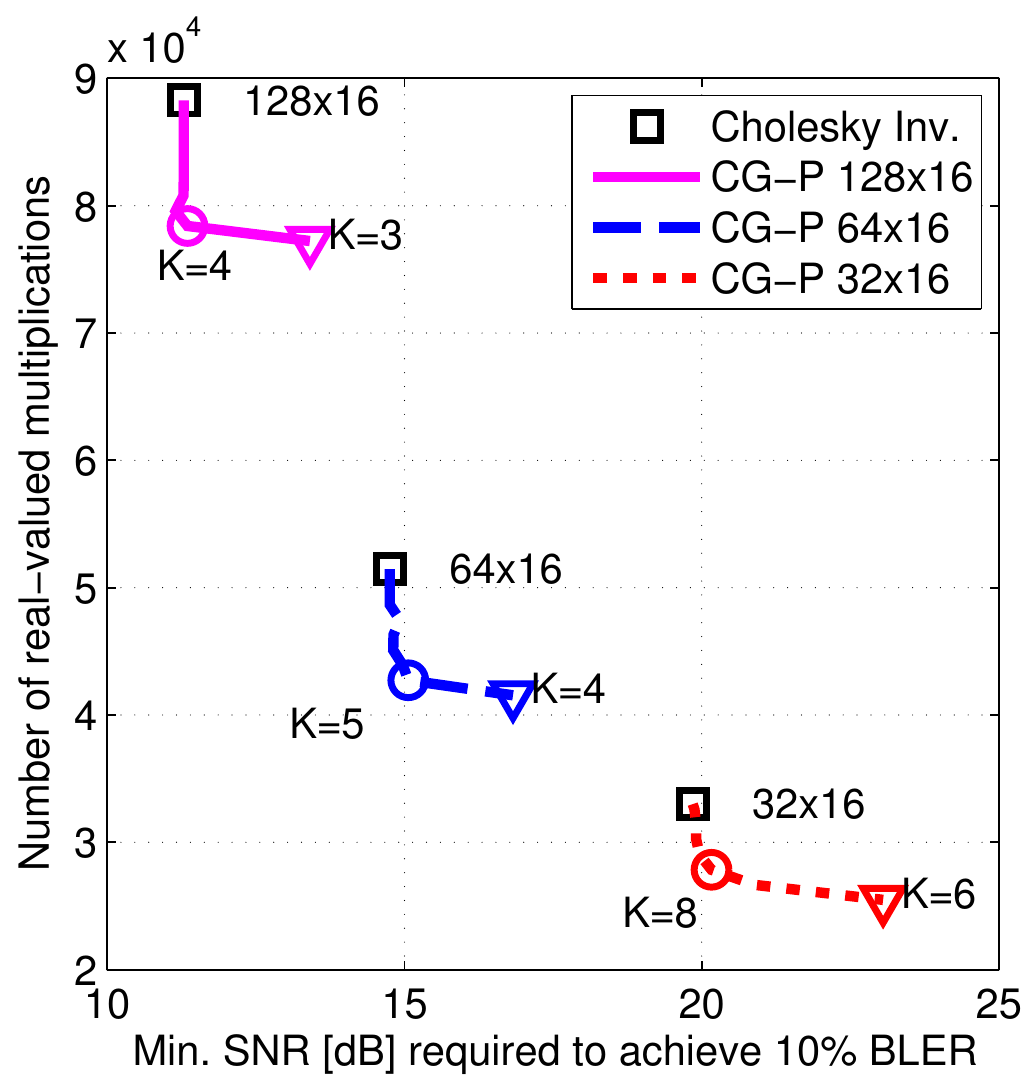}\label{fig:tradeoff64_16u_dl}}
\caption{Performance/complexity trade-off of the proposed CG-based precoding method  (CG-P) and the exact Cholesky-based precoder.}
\label{fig:tradeoff_dl}
\end{figure}

So far, we have only considered the uplink. In Figs.~\ref{fig:tradeoff_dl}(a)--(b) we briefly show the trade-offs achieved by CG-based and Cholesky-based precoding. 
We see that CG-based precoding is also able to achieve lower complexity at similar error-rate performance as the exact Cholesky-based precoder.

\section{Conclusions}
\label{sec:conclusions}
 
We have proposed a novel low-complexity  soft-output detection and precoding algorithm for the massive MIMO uplink and downlink, respectively. The proposed approach builds upon conjugate gradient (CG) methods and includes a novel post-equalization SINR tracking scheme, which is crucial to enable soft-output detection at low computational complexity. 
Our numerical results reveal that for reasonably large ratios between base station and user antennas, our CG-based detection and precoding approach quickly converges to that of an exact method. 
As a consequence, the proposed CG approach achieves an error-rate performance that is close to that of an exact inversion method, while requiring (often significantly) lower computational complexity. In addition, our CG-based scheme outperforms the approximate Neumann series inversion proposed in \cite{Wu2014} in terms of performance and complexity. In summary, the proposed approach is well suited for soft-output detection and precoding in realistic massive MIMO systems, and the algorithm's regularity and low complexity paves the way for efficient, high-throughput hardware designs.


\bibliographystyle{IEEEbib}
\bibliography{IEEEabrv,Massive_MIMO_bibfile}

\begin{thebibliography}{10}

\bibitem{Rusek2012}
F.~Rusek, D.~Persson, B.~K. Lau, E.~G. Larsson, T.~L. Marzetta, O.~Edfors, and
  F.~Tufvesson,
\newblock ``Scaling up {MIMO}: Opportunities and challenges with very large
  arrays,''
\newblock {\em IEEE Signal Process. Mag.}, vol. 30, no. 1, pp. 40--60, Jan.
  2013.

\bibitem{Marzetta2010}
T.~L. Marzetta,
\newblock ``Noncooperative cellular wireless with unlimited numbers of base
  station antennas,''
\newblock {\em IEEE Trans. Wireless Commun.}, vol. 9, no. 11, pp. 3590--3600,
  Nov. 2010.

\bibitem{Huh2011}
H.~Huh, G.~Caire, H.~C. Papadopoulos, and S.~A. Ramprashad,
\newblock ``Achieving ``massive {MIMO}'' spectral efficiency with a
  not-so-large number of antennas,''
\newblock {\em IEEE Trans. Wireless Commun.}, vol. 11, no. 9, pp. 3266--3239,
  Sept. 2012.

\bibitem{Ngo2012}
H.~Q. Ngo, E.~G. Larsson, and T.~L. Marzetta,
\newblock ``Energy and spectral efficiency of very large multiuser {MIMO}
  systems,''
\newblock {\em arXiv preprint: 1112.3810v2}, May 2012.

\bibitem{hoydis2011massive}
J.~Hoydis, S.~Ten~Brink, and M.~Debbah,
\newblock ``Massive {MIMO}: How many antennas do we need?,''
\newblock in {\em 49th Ann. Allerton Conf. on Commun., Control., and Comput.},
  Monticello, IL, Sept. 2011, pp. 545--550.

\bibitem{Wu2012}
M.~Wu, B.~Yin, A.~Vosoughi, C.~Studer, J.~R. Cavallaro, and C.~Dick,
\newblock ``Approximate matrix inversion for high-throughput data detection in
  the large-scale {MIMO} uplink,''
\newblock in {\em Proc. IEEE ISCAS}, Beijing, China, May 2013, pp. 2155--2158.

\bibitem{Yin2013}
B.~Yin, M.~Wu, C.~Studer, J.~R. Cavallaro, and C.~Dick,
\newblock ``Implementation trade-offs for linear detection in large-scale
  {MIMO} systems,''
\newblock in {\em Proc. IEEE ICASSP}, Vancouver, Canada, May 2013, pp.
  2679--2683.

\bibitem{Wu2014}
M.~Wu, B.~Yin, G.~Wang, C.~Dick, J.~R. Cavallaro, and C.~Studer,
\newblock ``Large-scale {MIMO} detection for {3GPP LTE:} algorithms and {FPGA}
  implementations,''
\newblock {\em IEEE J. Sel. Topics in Sig. Proc.}, 2014.

\bibitem{YWWDCS14b}
B.~Yin, M.~Wu, G.~Wang, C.~Dick, J.~R. Cavallaro, and C.~Studer,
\newblock ``A 3.8 {Gb/s} large-scale {MIMO} detector for {3GPP LTE-Advanced},''
\newblock in {\em Proc. IEEE ICASSP}, 2014.

\bibitem{winner2}
L.~Hentil{\"a}, P.~Ky{\"o}sti, M.~K{\"a}ske, M.~Narandzic, and M.~Alatossava,
\newblock ``Matlab implementation of the {WINNER} phase {II} channel model
  ver~1.1,'' Dec. 2007.

\bibitem{Paulraj2008}
A.~Paulraj, R.~Nabar, and D.~Gore,
\newblock {\em Introduction to Space-Time Wireless Communications},
\newblock Cambridge University Press, New York, USA, 2008.

\bibitem{Studer2011}
C.~Studer, S.~Fateh, and D.~Seethaler,
\newblock ``{ASIC} implementation of soft-input soft-output {MIMO} detection
  using {MMSE} parallel interference cancellation,''
\newblock {\em IEEE J. Solid-State Circuits}, vol. 46, no. 7, pp. 1754--1765,
  Jul. 2011.

\bibitem{Peel2005}
C.~Peel, B.~Hochwald, and A.~Swindlehurst,
\newblock ``A vector-perturbation technique for near-capacity multiantenna
  multiuser communication-part {I}: channel inversion and regularization,''
\newblock {\em IEEE Trans. Commun.}, vol. 53, no. 1, pp. 195--202, 2005.

\bibitem{Hestenes1952}
M.~R. Hestenes and E.~Stiefel,
\newblock ``Methods of conjugate gradients for solving linear systems,''
\newblock {\em J. Res. N.B.S. 49}, 1952.

\bibitem{Bulirsch91}
R.~Bulirsch and Stoer,
\newblock {\em Introduction to Numerical Analysis},
\newblock New York: Springer-Verlag, 1991.

\bibitem{Paige1982}
C.~Paige and M.~A. Saunders,
\newblock ``{LSQR}: An algorithm for sparse linear equations and sparse least
  squares,''
\newblock {\em ACM Trans. Math. Softw.}, vol. 8, pp. 43--71, 1982.

\bibitem{schreiber1986systolic}
R.~Schreiber and W.-P. Tang,
\newblock ``On systolic arrays for updating the {Cholesky} factorization,''
\newblock {\em BIT Numerical Mathematics}, vol. 26, no. 4, pp. 451--466, Dec.
  1986.

\end{thebibliography}

\end{document}